\documentclass[vecphys]{svmult}

\usepackage{makeidx}         % allows index generation
\usepackage{graphicx}        % standard LaTeX graphics tool
\usepackage{multicol}        % used for the two-column index
\usepackage[bottom]{footmisc}% places footnotes at page bottom
\makeindex             % used for the subject index
\begin{document}

\title*{The Effelsberg--Bonn HI Survey (EBHIS)}
\author{B. Winkel\inst{1,2}\and
J. Kerp\inst{2} \and P. Kalberla\inst{2} \and R. Keller\inst{1}}
\institute{Max-Planck-Institut f\"{u}r Radioastronomie, International Max Planck Research School (IMPRS) for Radio and Infrared Astronomy at the Universities of Bonn and Cologne, Auf dem H\"{u}gel 69, 53121 Bonn, Germany \texttt{bwinkel@astro.uni-bonn.de}
\and Argelander-Institut f\"{u}r Astronomie, Auf dem H\"{u}gel 71, 53121 Bonn, Germany
}
\maketitle
The new L-band 7-feed-array at the 100-m telescope in Effelsberg will be used to perform an unbiased fully sampled HI survey of the entire northern hemisphere observing the galactic and extragalactic sky using simultaneously two different backends. The integration time per position will be 10\,min towards the SDSS area and 2\,min for the remaining sky, thereby achieving a sensitivity competitive with the Arecibo ALFALFA and GALFA surveys but covering a much larger area of the sky.

Both backends are FPGA-based digital fast fourier transform (DFFT) spectrometers, offering a superior dynamic range and temporal resolution. The latter is crucial for a sophisticated RFI mitigation scheme (Winkel et al. 2007) but produces huge amounts of data over the projected five years of observing. Consequently, we put considerable effort into efficient data reduction. Table\,\ref{tabsurveyparameters} compares the survey parameters with several recent HI surveys. %The Effelsberg HI Survey (EBHIS) is designed to compete in all respects with them.

%\centering
\begin{table}
\caption{Parameters of several recent HI surveys in compared to EBHIS. }
\label{tabsurveyparameters}		
\begin{tabular}{cccccccc}
\hline\noalign{\smallskip}
%\small
Survey	&Telescope		&Type	&Status		&Area			&Beam size	&$z_\mathrm{max}$	&$\Delta v$			\\		%&Integr. time	
	&			&	&	 	&(sq.deg.)		&(arcmin)	&			&(km/s)				\\%\hline	%& per beam	
\noalign{\smallskip}\hline\noalign{\smallskip}
 LAB	&Dwing/Arg		&g	&Completed	&41\,300		&36		&			&1.3				\\		%&		
HIPASS	&Parkes			&e	&Completed	&29\,300		&14.1		&~0.05			&18				\\		%&450		
GASS	&			&g	&Ongoing	&20\,600		&		&			&0.8				\\		%&90		
ALFALFA	&Arecibo		&e	&Ongoing	&7\,100			&3.4		&~0.06			&11				\\		%&28		
GALFA	&			&g	&Ongoing	&			&		&			&0.7				\\		%&18		
EBHISe	&Effelsberg		&e	&Planned	&20\,600 (8\,500$^\ast$)&9		&~0.07			&7				\\		%&120(600$^\ast$)
EBHISg	&			&g	&Planned	&			&		&			&1				\\%\hline	%&
\end{tabular}

\begin{tabular}{cccccc}%\hline
\hline\noalign{\smallskip}
Survey	&RMS noise	&$N_\mathrm{HI}$ limit		&Mass limit			&Velocities					&Integr. time\\		%&Integr. time	
	&(mJy/Beam)	&($10^{18}\,\mathrm{cm}^{-2}$)	&($10^7\,\mathrm{M}_\odot$)	&(km/s)	   					& per beam (s)			\\%\hline	%& per beam	
\noalign{\smallskip}
\hline\noalign{\smallskip}
 LAB	&620	   	&3.3    			&		   		&$-\phantom{1\,}450\ldots \phantom{10\,}400$   	& \\		%&		
HIPASS	&13	   	&	   			&8.7 	   			&$-1\,280\ldots 12\,700$ 			&450\\		%&450		
GASS	&95	   	&2.6    			&		   		&$-\phantom{1\,}400\ldots \phantom{10\,}450$   	&90 \\		%&90		
ALFALFA	&1.6	   	&	   			&0.8 	   			&$-2\,000\ldots 18\,000$ 			&28\\		%&28		
GALFA	&13.2	   	&3.2    			&		   		&$-\phantom{1\,}700\ldots \phantom{10\,}700$    &18\\		%&18		
EBHISe	&4.5 (2$^\ast$) &	   			&1.9 (0.8$^\ast$)    		&$-2\,000\ldots 18\,000$ 			&120(600$^\ast$)\\		%&120(600$^\ast$)
EBHISg	&12 (5.5$^\ast$)&0.9 (0.4$^\ast$)		&  	   			&$-\phantom{1\,}500\ldots \phantom{10\,}500$	&\\%\hline	%&
\noalign{\smallskip}\hline\noalign{\smallskip}
\end{tabular}
$^\ast$Towards the SDSS area EBHIS has higher sensitivity.
\end{table}

The data processing is organized making use of individual data reduction modules. Each module calculates correction factors/spectra that are stored in a database. A special merger module is used to apply all corrections to the data before the gridder produces a final data cube. 
%Accordingly, it is feasible to produce data cubes i.e. corrected for RFI events but without stray-radiation correction or vise versa. 
The advantage of this approach is the flexibility to modify individual modules without having to recompute all  corrections.

Most of the reduction modules are ready-for-use. 
%Many modules are optimized to be performed on multi-processor computers. 
Winkel et al. (2007) describe a very sensitive RFI detection scheme, utilizing the high temporal resolution spectra from the DFFT spectrometers. The stray-radiation correction is performed using the method of Kalberla et al. (2005). Their approach enables stray radiation correction even within the regime of the high-velocity clouds. 
%As an example, the Magellanic Stream data needs to be corrected for stray-radiation towards the southern galactic pole. 
We also implemented and improved (see Winkel \& Kerp 2007) the recently proposed least-squares frequency switching (LSFS) method of Heiles (2007). Using more than two different intermediate frequency settings, one is able to calculate the baseline even for faintest HI emission with unprecedented quality.
Finally, we developed a graphical user interface (GUI) optimized for the search and parametrization of galaxies in the HI data cubes. We implemented a very promising finder algorithm based on the Gamma test (Boyce 2003). The GUI is designed to allow a fast working flow and is able to compute statistical errors using Markov chain Monte Carlo methods.

EBHIS will be extremely valuable for a broad range of research topics: study of the low-mass end of the HI mass function (HIMF) in the local volume, environmental and evolutionary effects (as seen in the HIMF), the search for galaxies near low-redshift Lyman-alpha absorbers, and analysis of multiphase and extraplanar gas, HI shells, and ultra-compact high-velocity-clouds (Br\"{u}ns \& Westmeier 2004).

\printindex
\end{document}